# A New Hyperchaotic Attractor with Complex Patterns

**Safieddine Bouali**
University of Tunis, Management Institute,
Department of Quantitative Methods & Economics,
41, rue de la Liberté, 2000, Le Bardo, Tunisia
Safieddine.Bouali@isg.rnu.tn

*The paper introduces a new dynamical system which equation's specification ensures full hyperchaotic attractor patterns. Its focal statement appears in its self-sufficiency from previous 3D nonlinear systems. It is mathematically not an extension of a 3D nonlinear system into the fourth dimension but it is built as a new system integrating a small set of nonlinear terms. To explore the basic behavior of the new 4D dynamical model, we selected an arrangement of simple parameters to display the phase portraits of the related strange attractor projected onto the 3D representation spaces.*

*The computation of the Lyapunov exponents establishes presence of hyperchaos since two positive exponents are found. Indexes of stability of the equilibrium points corresponding to the typical 4D strange attractor are also examined. A collection of Poincaré Maps highlights the intricate dynamics of the system.*

# 1. Introduction

Hyperchaos concept was firstly introduced in the seminal paper of Rössler [1] to assert the dynamical patterns of dynamical systems when more than one positive Lyapunov exponent is found [2]. Some of hyperchaotic systems are thereafter discovered mainly by the extension of established 3D chaotic to the four-dimension. These hyperchaotic applications start from a fully 3d chaotic system adding a state feedback controller [see for example, 3-9]. Such 4D models are only expanded editions in the phase hyperspace.

This paper introduces a new hyperchaotic system not derived from previous 3d models since we do not apply *hyperchaotification* techniques. Indeed, all the 3D sub-models of the system are not sustainable. However, the present system retains a modified 2D Lotka-Volterra oscillator [10-11] integrated in the block of the first two equations.

Purposefully creating hyperchaos can be a nontrivial task to focus a new kind of dynamical patterns as shown by several 4D models recently discovered [12-18, and references therein].

Section 2 investigates the basic characteristics of the introduced autonomous four-dimensional system of first order differential equations. Section 3 points chiefly to the intricate patterns of the hyperchaotic attractor generated by the system. Concluding remarks reports the singularity of the model and its novel contribution to the chaos literature.

# 2. The hyperchaotic 4D model

The new system is governed by four nonlinear differential equations:

$$\begin{cases} dx/dt = x(1-y) + \alpha z \\ dy/dt = \beta(x^2 - 1) y \\ dz/dt = \gamma(1-y) v \\ dv/dt = \eta z \end{cases} \quad (1)$$

where x, y, z and v the state variables of the model, and α, β, γ, and η real parameters.

Equations embed only height terms on the right-hand side, three of them are nonlinear, i.e. two quadratic terms *xy*, and vy, and a unique cubic cross-term $yx^2$.

One would expect the emergence of hyperchaotic patterns in our application.

We notice that the system is not derived from a previous 3d chaotic model. Any configuration of systems carried out by three state variables amongst the four is not sustainable.
Indeed, the introduced 4D model is not an expanded version adding a state feedback loop formulated in a supplementary equation.
In fact, the core of the system is constituted by the 2D sub-model representing a modified version of the well-established 2D Lotka-Volterra oscillator,

$$\begin{cases} dx/dt = x(1-y) \\ dy/dt = (x^2 - 1) y \end{cases}$$

and composing previously the core of typical three dimensional chaotic systems. Indeed, this basic oscillator was applied as the fundamental mechanism of two chaotic models [19-20].
The exploration of the widest dynamical behaviors leads us to choose specific values of the parameters between several specifications to simplify the analysis.
Let $P_0$ ($\alpha$, $\beta$, $\gamma$, $\eta$) = (-2, 1, 0.2, 1), the equilibria of the system (1) are found by setting $dx/dt = dy/dt = dz/dt = dv/dt = 0$, expressed as follow:

$$\begin{cases} 0 = x(1-y) - 2z \\ 0 = (x^2 - 1) y \\ 0 = 0.2 (1-y) v \\ 0 = z \end{cases} \quad (1.1)$$

The coordinates of the equilibria are the origin $S_0$, (0, 0, 0, 0) and two sets of parametric solutions: $S_1$ (1, 1, 0, v) and $S_2$ (-1, 1, 0, v).
The eigenvalues $\lambda_i$ and stability features of the related solutions are determined from the characteristic equation $|J - \lambda I| = 0$, where $I$, the unit matrix and $J$, the Jacobian of the model:

$$J = \begin{vmatrix} 1-y & -x & -2 & 0 \\ 2xy & (x^2-1) & 0 & 0 \\ 0 & -0.2v & 0 & 0.2(1-y) \\ 0 & 0 & 1 & 0 \end{vmatrix}$$

The volume contraction of the flot is given by: $\nabla V = Tr(J) = x^2 - y.$

However, the dissipativity in the phase hyperspace is enclosed in the domain: $x^2 - y < 0$, and thus the system is a dissipative system. In this domain, the orbits converge to a specific subset of zero volume as $t \to \infty$ exponentially, i.e. $dV/dt = e^{(x^2 - y)}$.

Indeed, for $P_0$, all solutions are unstable (Table 1) and distinguished with different indexes of stability.

| Coordinates of the Equilibrium | The corresponding characteristic equation, $|J - \lambda I| = 0$, and eigenvalues | Index [1] and Instability |
|---|---|---|
| $S_0 (x_0, y_0, z_0, v_0) = (0, 0, 0, 0)$<br><br>Det $(J_0) = 0.2 > 0$ | $(\lambda^2 - 1)(\lambda^2 - 0.2) = 0$<br>$\lambda_1 = -1 \quad \lambda_2 = -0.2$<br>$\lambda_3 = 0.2 \quad \lambda_4 = 1$ | **Index-2:** *saddle focus* |
| Set $S_1$ of equilibria:<br>$S_1 (x_1, y_1, z_1, v_1) = (1, 1, 0, v)$<br>for $v \in R$<br><br>Det $(J_1) = 0$ | *For $V \geq 0$* $\quad$ Re $(\lambda_1) > 0$<br>$\qquad\qquad$ Re $(\lambda_2, \lambda_3, \lambda_4) \leq 0$ | **Index-1** *Saddle points* |
| | *iFor $V < 0$* $\quad$ Re $(\lambda_1, \lambda_2) > 0$<br>$\qquad\qquad$ Re $(\lambda_3, \lambda_4) \leq 0$ | **Index-2:** *saddle foci* |
| Set $S_2$ of equilibria:<br>$S_2 (x_2, y_2, z_2, v_2) = (-1, 1, 0, v)$<br>for $v \in R$<br><br>Det $(J_2) = 0$ | *For $V > 0$* $\quad$ Re $(\lambda_1, \lambda_2) > 0$<br>$\qquad\qquad$ Re $(\lambda_3, \lambda_4) \leq 0$ | **Index-2:** *saddle foci* |
| | *For $V \leq 0$* $\quad$ Re $(\lambda_1) > 0$<br>$\qquad\qquad$ Re $(\lambda_2, \lambda_3, \lambda_4) \leq 0$ | **Index-1** *Saddle points* |

(1) Index reports the number of eigenvalues with real parts Re $(\lambda) > 0$. From 1 to 4, it indicates the degree of instability. Index-0: null or negative real parts of all eigenvalues of the equilibrium characterize its stability.

**Table 1. Stability of the Equilibria**

## 3. Global Behavior of the hyperchaotic Attractor

To substitute the unfeasibility of the four-dimensional representation, four projections in 3D phase portraits are exhibited, respectively the x-y-z, the x-y-v, the z-y-v, and z-v-x spaces (Fig. 1). The orbits have intricate paths following butterfly wings and scrolls. The trajectories are extremely abundant and dense taking relatively regular and simple forms. However, does the new system have explicitly and globally a hyperchaotic nature?

It is known that the spectrum of Lyapunov exponents is the most useful diagnostic to quantify chaos. When the nearby trajectories in the phase space diverge at exponential rates, giving positive Lyapunov exponents, the dynamics become unpredictable. Any system containing at least two positive Lyapunov exponents is defined to be hyperchaotic.

The computed Lyapunov exponents $LE_i$ are the follows:

$LE_1 = 1.85$ $\qquad LE_2 = .36 \qquad LE_3 \approx 0 \qquad LE_4 = -34.2$

The system (1) can be classified *hyperchaotic* since: $LE_1 > LE_2 > 0$, $LE_3 = 0$, $LE_4 < 0$ and $LE_1 + LE_2 + LE_4 < 0$.

The Kaplan-Yorke dimension of the attractor reaches D $\approx$ **3.064.**

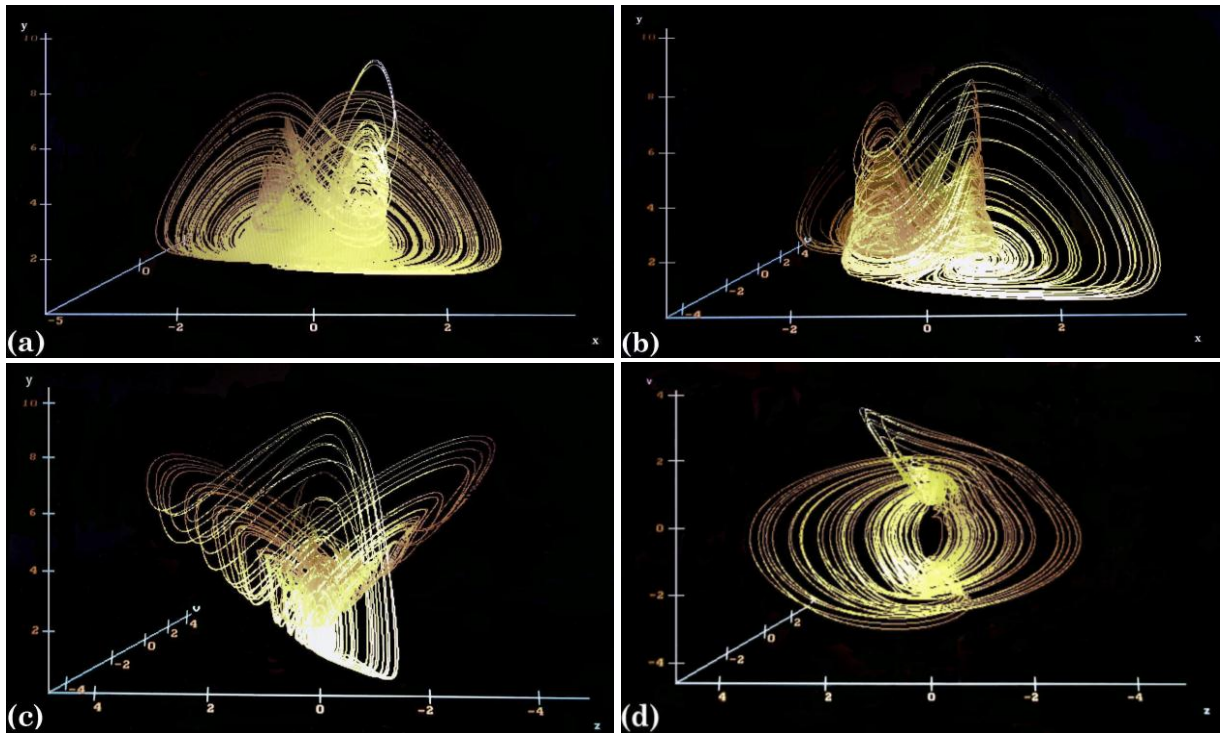

**Fig.1. Projections of the 4D hyperchaotic system into the 3D phase portraits.**

(**a**) Projection on x-y-z space, (**b**) Projection on x-y-v space , (**c**) Projection on z-y-v space, and (**d**) Projection on z-v-x space

To describe the folding properties of chaos, we apply the technique of the Poincaré map. Take $\sum = \{(x, y, z, v) \in R^4 \mid y = 2\}$ as a crossing section. We select the mapping on $\sum$ from the 4D space not to the 3D phase space but on 2D surfaces.

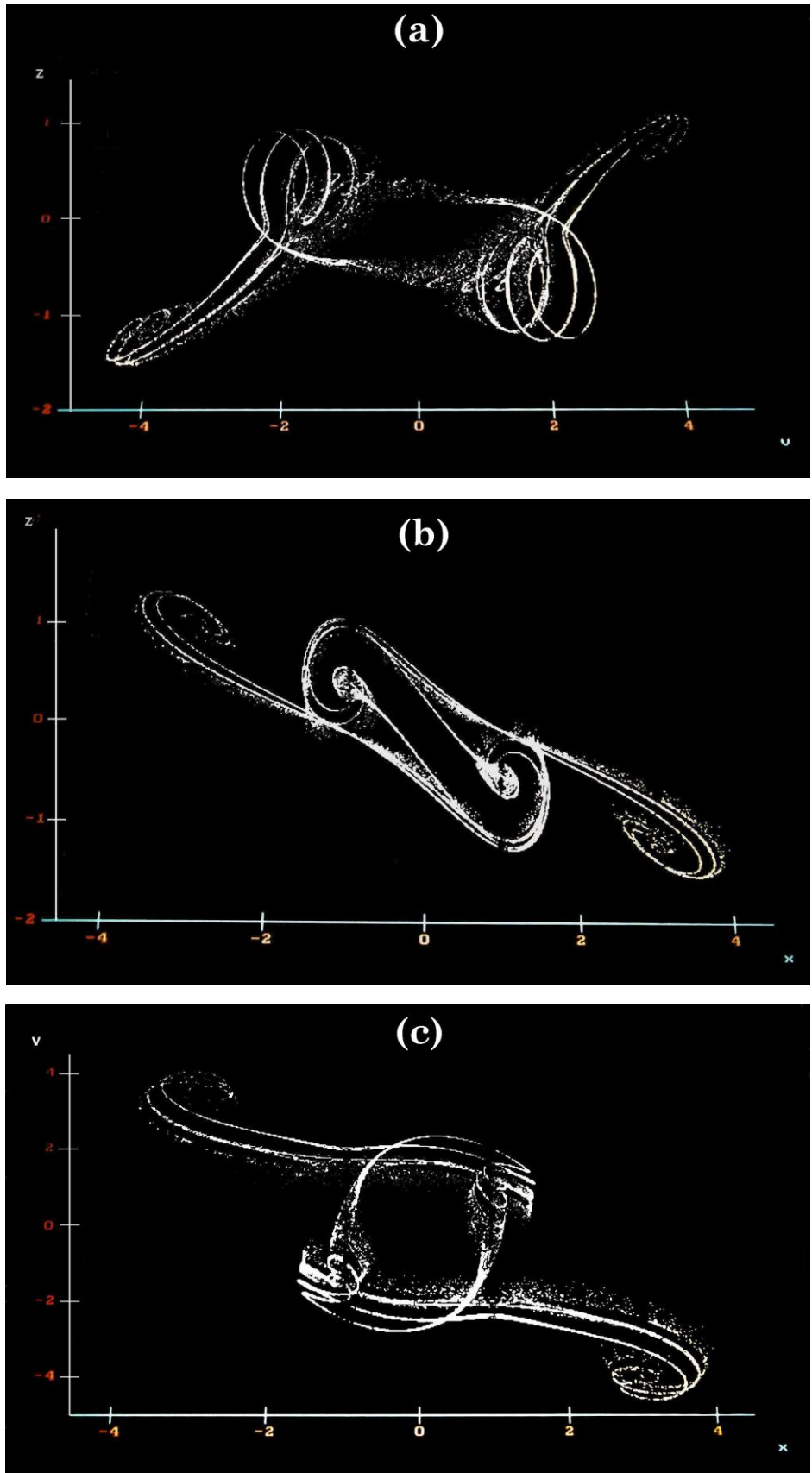

**Fig.2. Poincaré maps of the System (1) for Y = 2.**

**(a)** Projection on *v*–z of the Poincaré map, **(b)** Projection on x–z of the Poincaré map, and **(c)** Projection on *x*–v of the Poincaré map.

The associated Poincaré map on the three phase planes show that the system has extremely rich and intricate dynamics (fig. 2). Branches and twigs unravel the complex folding of the hyperchaotic attractor. In fact, such mapping can exhibit nine other 2D maps for x, z and v as crossing sections. Similar dynamical patterns are also found indicating complex envelops of the orbits.

## 4. Concluding Remarks

The prevailing feature of the system emerges in its independency from previous 4D chaotic systems. Reaching chaos through a relatively simple algebraic structure is still a challenging task. The new hyperchaotic system embedding a modified 2D Lotka-Volterra oscillator depicts a complex scroll butterfly-shaped attractor and exhibiting elegant contours.

On the other hand, several specifications of P have been experimented. The simplest one, $P_0$, inducing the widest range of dynamical behaviors, had been selected.

This intentionally constructed 4d chaotic system doesn't reincarnate known hyperchaotic patterns. The appearance and also the characteristics of the new 4D attractor are utterly distinct from the other existing hyperchaotic systems (4D Lorenz–Haken system, 4D hyperchaotic Chua's circuit, 4D hyperchaotic Chen, etc.).

The new system could be suitable for the financial modelization of the macroeconomic sphere. It could be also suitable for digital signal encryption in the communication field when its variants provide a very large set of encryption keys. Further developments and extended analysis related to the set of parameters, and the diagram of bifurcations, will be investigated in a future work.

**March 2015**